\documentclass[reprint,twocolumn]{revtex4-1}
\usepackage{graphicx}
\usepackage{dcolumn}%
\usepackage{bm}%
\usepackage{amssymb}
\usepackage{nameref}
\usepackage{epsfig,colordvi}
\usepackage[latin1]{inputenc}
\usepackage{color} 
\usepackage{here}

 \def\be{\begin{equation}}
 \def\ee{\end{equation}}
 \def\bea{\begin{eqnarray}}
 \def\eea{\end{eqnarray}}
 \def\bean{\begin{eqnarray*}}
 \def\eean{\end{eqnarray*}}
 \def\gsim{\mathrel{\rlap{\lower0.2em\hbox{$\sim$}}\raise0.2em\hbox{$>$}}}
 \def\ksim{\mathrel{\rlap{\lower0.2em\hbox{$\sim$}}\raise0.2em\hbox{$<$}}}
 \def\kg{\mathrel{\rlap{\lower0.25em\hbox{$>$}}\raise0.25em\hbox{$<$}}}

\newcommand{\AuAu} {${\mbox{$^{197}Au$}}+{\mbox{$^{197}Au$}}$ }
\newcommand{\XeSn} {$\mbox{$^{129}Xe$}+\mbox{$^{124}Sn$}$ }
\newcommand{\LiC} {$\mbox{$^{6}Li$}+\mbox{$^{12}C$}$ }

\begin{document}

\title{FRIGA, A New Approach To Identify Isotopes and  Hyper-nuclei In N-Body Transport Models.}

\author{A. Le F\`evre$^1$, J. Aichelin$^{2,3}$, C. Hartnack$^2$, Y. Leifels$^1$}
\affiliation{$^1$ GSI Helmholtzzentrum f\"ur Schwerionenforschung GmbH,
  Planckstr. 1, 64291 Darmstadt, Germany} 
\affiliation{$^2$ SUBATECH, IMT Atlantique, Universit\'e de Nantes, IN2P3/CNRS
\\ 4 rue Alfred Kastler, 44307 Nantes cedex 3, France}
\affiliation{$^3$ Frankfurt Institute for Advanced Studies, Ruth Moufang Str. 1
\\ 60438 Frankfurt, Germany}

\date{\today}

\begin{abstract} \noindent
We present a new approach to identify fragments in computer simulations of
relativistic heavy ion collisions. It is based on the simulated annealing
technique and can be applied to n-body transport models like the  Quantum
Molecular Dynamics. This new approach is able to predict isotope yields as
well as hyper-nucleus production. In order to illustrate its predicting power,
we confront this new method with experimental data and show the sensitivity on
the parameters which govern the cluster formation. 
\end{abstract}


\maketitle

\section{Introduction}
In heavy ion reactions at energies between 20~MeV and several GeV per nucleon, 
the formation of complex clusters is a key observable \cite{sch96,rei10}, which is not
understood in all details, yet. Sophisticated microscopic transport models \cite{Feldmeier90,Ono92,Danielewicz01} employing clusters as
degrees of freedom have been developed, but they are not
generally applicable to the collision energies under consideration or
are constrained to small clusters (A $\leq$ 3). 
Identifying clusters
represents also a major challenge for transport
models propagating only nucleons as relevant degrees of freedom.
Omitting fragment formation makes the prediction of proton and neutron
observables ambiguous because fragments have different kinematical properties
than single nucleons. 

Transport models based on
the time evolution of the one-body density matrix, like BUU \cite{Buss:2011mx}
or SMASH \cite{Weil:2016zrk} cannot address this question directly without
injecting sufficient phase space fluctuations into the system \cite{Bauer87,Guarnera96,Colonna98}.   

On their side, QMD approaches, which are based on N-body theories \cite{aic91,har98,Bass:1998ca}
propagate the correlations in time and, therefore, contain all necessary
information to describe clusters.   

Basic ways to identify clusters are to employ a coalescence model
\cite{gut76,Gosset77,lem79,sat81} or to use a minimum spanning tree (MST)
procedure \cite{gos97}. The first method needs 
various coalescence parameters for each isotope, is unable to deal with heavy
clusters, and moreover a time has to be chosen at which the transport
calculation is stopped and the coalescence
procedure is applied.  In a rapidly expanding system 
like at the end of a heavy ion collision the cluster yields depend crucially
on this time. 
In addition, it has been shown for light clusters, that this
time  is different for different  isotopes \cite{Gossiaux:1994jq}. To
study the production of light hypernuclei in the UrQMD model the coalescence
model has recently been applied in
refs. \cite{Steinheimer:2012tb,Botvina:2016fav}. It turns out that there
exists a choice of parameters for which the results are quite satisfying. The
MST procedure -- based only on proximity criteria in position and optionally momentum space
-- allows only for an identification of the fragments at the end of the
reaction when the fragments and single nucleons  have well separated from each
other. 

The drawback of both methods is that the study of the
physical origin \cite{gos97} of fragmentation is excluded. In addition, since the
underlying transport models are semi-classical, binding energy modifications
due to closed shells or pairing energies are neglected in these approaches.

\section{The principles of the fragment recognition.}
Identifying fragments early, while the reaction is still going on,
requires to define the most bound cluster partition
out of a set of clusters pre-selected using momentum as well as coordinate
space information, like in MST.
This idea has been first introduced by Dorso et al. \cite{dor93}. It has
been further applied into the 
Simulated Annealing Clusterisation Algorithm (SACA) \cite{pur00} in the late
1990's and has been successfully applied to  
understand experimental fragment charge distribution and spectra as well as
bimodal distributions \cite{gos97, zbi07, lef09}. This procedure can be applied at
different times during the collision to provide the time evolution of the most
probable cluster distribution in terms of binding energy. It turns out that
this method detects the final partitions early -- that found by MST at very late times (typically $>200$~fm/c) --, 
right after the colliding
system begins to separate, when the
energetic collisions are over, as shown in \cite{pur00}. 
By ''final partition'', we mean the asymptotic (late time) cluster distribution that would be detected by the same method within a 
transport model that would not introduce artificial time instabilities in the phase space distribution of nucleons.
We want to stress at this point the fact that, when we are using for example a QMD transport model, we are dealing with a semi-classical approach, 
and not with a pure-quantum approach. This manifests itself on a long time scale when the fragments become unstable. 
The reason for this is that in these codes the ground state of the quantum hamiltonian is higher than the ground state of the classical hamiltonian. 
As a consequence, a semi-classical system can still emit particle when in the analog quantum system it is not possible anymore.   

The seed, an ensemble of  pre-clusters, for the annealing procedure is generated by using an MST. 
Then nucleons are exchanged between neighbouring fragments or single
nucleons in all possible
ways applying a simulated annealing technique, based on a Metropolis
algorithm. Neglecting the interactions between nucleons of different clusters,
but taking into account the interaction  
among the nucleons in the same fragment, this algorithm identifies the
combination of fragments and free nucleons 
which has the \textit{most negative total binding energy} -- i.e. the most bound sum.
The reason for this is the fact that fragments
are not a random collection of nucleons  at the end, but initial-final
state correlations. 

In SACA the nucleon-nucleon interactions taken into account to calculate the
binding energies of clusters 
are a Skyrme potential supplemented by a Yukawa term -- QMD surface
correction \cite{aic91} -- and a Coulomb potential.  
These potentials are also used for the propagation of the nucleons in the QMD
transport model which was utilised for the time evolution 
of the reaction \cite{aic91}.
They are used for calculating a cluster binding energy in the following way:
introducing (density dependent) two-body interactions among the nucleons
which form a fragment,  
the internal fragment energy
\begin{eqnarray}
E_B &=& \langle H \rangle = \langle T \rangle + \langle V \rangle(N,Z)
\nonumber \\ 
&=& \sum_i \frac{p_i^2}{2m_i} +
\nonumber \\ 
\sum_{i} \sum_{j>i}
 \int f_i({\bf r, p},t) \,
&V&({\bf r,r\,',p,p\,'})  f_j({\bf r\,',p\,'},t)\, 
\nonumber \\ 
&&\rm d{\bf r}\, \rm d{\bf r\,'}\rm d{\bf p}\, \rm d{\bf p\,'} \quad. 
\label{binding}
\end{eqnarray}
where {\bf r, p} is the particle phase-space position in the centre-of-mass of
the collision,  $m_i$ is its mass,  $V$ is the potential, $f_i$ is the
single-particle Wigner 
density 
\begin{equation} \label{fdefinition}
 f_i ({\bf r, p},t) = \frac{1}{\pi^3 \hbar^3 }
 {\rm e}^{-\frac{2}{L} ({\bf r} - {\bf r_i} (t) )^2   }
 {\rm e}^{-\frac{L}{2\hbar^2} ({\bf p - p_i} (t) )^2 } \quad 
\end{equation}

The potential $V$ consists out of a Skyrme type potential
complemented by a Yukawa and a Coulomb potential. This
combination of potentials we will denote as ''basic''.   
It has been shown in reference \cite{aic91}
that Eq.~\ref{binding} reproduces very well the binding energies of nuclei 
with $A>5$ as given by the Bethe-Weizs\"acker mass formula for ground state
nuclei, $B_{BW,0}$ (see Fig.12 of Ref. \cite{aic91}).
For nuclei with $A\leq5$ this method provides slightly less
bound values than the Bethe-Weizs\"acker mass formula. This is taken into
account in FRIGA by 
shifting accordingly the cluster ground state binding energy when calculating its excitation energy. 
Note that the nuclear densities in a cluster are here computed from the sole nucleons composing it, as if it were isolated, similarly to neglecting the nuclear force from external nucleons. 
The reason is that as soon as the asymptotic cluster partition is reached, the cluster binding energy must correspond to that of a free nucleus, therefore whose 
relevant density is that of its asymptotic state, i.e. close to the ground state. For this reason, the density used for calculating the cluster binding energy differs from that of the medium. 

The SACA model has been extended in order to predict more
realistically fragment yields in the isotopic degree of freedom, and to address 
hyper-nuclei production. For doing so, additional potentials (asymmetry energy
and shell effects) enter the determination of the binding energy of primary
clusters. In addition, when excited, those latter undergo a sequential
secondary decay at very late times, when the long range Coulomb interaction
between clusters becomes negligible (at the order of 1000 fm/c). 
This new approach is dubbed FRIGA ("Fragment Recognition In General Application"). 
The basic idea of FRIGA is to use the same potentials (mean
fields) as has been applied in the transport code and to add further interactions
which are relevant for binding energies of nuclei, in particular shell effects.    


\section{The features of FRIGA.}

In order to predict the isotope yields, we have extended the SACA cluster
identification algorithm by including asymmetry energy, pairing and 
shell effects.

For the asymmetry energy we adopt the parametrisation from  IQMD \cite{har98},  
the transport code which we use -- in addition to BQMD -- in the present
article for the transport of the 
nucleons. The potential part of the asymmetry energy, which is repulsive, thus reads:
\begin{displaymath}
  B_{asy}=E_{0} (\frac{\rho_{n}-\rho_{p}}{\rho_{B}})^{2} (\frac{<\rho_{B}>}{\rho_{0}})^{\gamma}
\end{displaymath}
where $E_{0}$=23.3 MeV, and
$\rho_{n}$, $\rho_{p}$, $\rho_{B}$, $\rho_{0}$ are the
neutron, proton, baryonic and saturation densities, respectively. In the
present work, by default, we take $\gamma$=1 (linear density dependence).\\ 
Note that the kinetic part of the asymmetry energy is carried by the
nucleon momenta which, according to the Thomas-Fermi model, relates to the
differences of the Fermi edges of neutrons and protons: For $\rho_0$ it
corresponds to a value of about 9 MeV in IQMD.


Another significant part of the binding energy of light
isotopes are the shell structure and odd-even effects (pairing). In the
conditions of high pressure and temperature where FRIGA is used to
determine the pre-fragments, these structure effects are not well
known. E. Khan et al. \cite{kha07} showed that there are some indications 
that they can affect the primary fragments. The authors
demonstrate that the pairing vanishes above a nuclear temperature $T_V\approx0.5\Delta_{pairing}$
 (pairing energy). 
 At the density of their fundamental state, 
 the pairing energy tends to be
 negligible for heavy nuclei, with the pairing energy taken from the Bethe-Weizs\"acker mass formula $\Delta_{pairing} = 11.2
 A^{-\frac{1}{2}}$~MeV (positive for even-even and negative for  odd-odd nuclei),
 whereas it is strong for light isotopes, like $^{4}He$ and $^{3}He$
 with 12~MeV and 6.9~MeV, respectively. In FRIGA, due to the minimisation of
 the binding energy, the primary fragments 
 are expected to be produced quite cold on average, with $T\sim1-2 MeV$, and with a density 
 close to that of their ground state, slightly below $\rho_0$  (typically between $\rho_0/2$ and $\rho_0$ depending on the fragment size). 
 Hence, their temperature could be
 below $T_V$ and one cannot neglect the pairing energy. The same 
 might be true for shell effects 
which produce 
a visible enhancement of the measured fragment yields for
closed shell nuclei. It will be a crucial point to determine whether  
these shell effects are already realised in the primary stage of the fragment
production, or later due to secondary de-excitation.  

In order to determine the contribution of all structure effects to the binding energy of primary
clusters 
identified by FRIGA, we make two hypotheses, independent of the
density and the average kinetic energy of the fragment environment. 

{\it First}, the relative ratio of the nuclear structure contribution to the
overall binding energy remains unchanged at 
moderate temperatures and at
densities close to that of the fundamental state of the cluster. 


Applying our first assumption that the ratio of the still ''unknown'' nuclear
structure contribution to the binding energy $B_{struct}(Z,N,\rho,T)$ and the
calculated binding energy $E_B(Z,N,\rho,T)$ of Eq.~\ref{binding} is constant 
in the respective density and temperature ranges, 
one obtains 
\begin{eqnarray}\label{structratio}
\frac{B_{struct}(Z,N,\rho,T)}{E_{B(Sky,Yuk,Coul)}(Z,N,\rho,T)} & = & \nonumber
\\
\frac{B_{exp,struct}(Z,N)}{B_{BW(vol,surf,Coul)}(Z,N)} & = & const.(Z,N)
\end{eqnarray}
where $B_{BW(vol,surf,Coul)}$ is the binding energy as given by the
sum of the volume, surface, and Coulomb terms of the
Bethe-Weizs\"acker mass formula -- considered as a ground state for $E_{B(Sky,Yuk,Coul)}(Z,N,\rho=rho_0,T=0)$, 
whereas  $B_{exp,struct}$ is the difference between the experimentally observed 
binding energy $B_{exp}$ and the prediction of the mass formula without pairing term, 
$B_{exp,struct}(Z,N)= B_{exp}(Z,N) - B_{BW(vol,surf,Coul,asy)}(Z,N)$. 

Our {\it second} hypothesis is that Eq.~\ref{binding} remains the correct
description of the binding energy if the nuclei are deformed or excited as it
might happen for fragments identified by the FRIGA algorithm.  

Under these assumptions we can express  the nuclear structure contribution to the binding energy
 of a deformed cluster with Z protons and N neutrons  in the following way:
\begin{eqnarray}
 B_{struct} (\rho,T,Z,N) & = & (B_{exp}(Z,N) - B_{BW np}(Z,N)) \nonumber 
 \\
 & \times & \frac{E_{B}(\rho, T, Z,N)}{B_{BW np, na}(Z,N)} \nonumber 
 \end{eqnarray} 
$E_B$ is the binding energy of Eq. \ref{binding} and $B_{BW np, na}$ is that given by the 
Bethe-Weizs\"acker formula binding without asymmetry ("na") and pairing ("np")
contributions. 
Isotopes and hyper nuclei which are not stable at all in nature, are discarded in FRIGA
by assigning to them a very 
repulsive $E_{B}$.
The complete total binding energy of a cluster with N and Z, which is used in the
annealing algorithm, will then be:  
\begin{displaymath}
 B = E_B(Z,N) + B_{asy} + B_{struct}.
 \end{displaymath}
in contradistinction to SACA in which only the first term is used. 

The other new feature of FRIGA concerns the initial configuration of the SACA
algorithm. There the simulated annealing procedure started out from the full
cluster partition provided by the MST procedure, based on the distance of the
nucleons in coordinate space. Subsequently  SACA reorganises the partition in
order to minimize the sum of the cluster binding energies. In FRIGA, "cold"
MST primary clusters are removed 
from the ensemble, i.e.are kept as they are. The standard criterium for
categorizing a fragment as  "cold"  is a 
maximum internal excitation energy of 1A~MeV (see definition in
Sect.~\ref{sec:second}). This new method is particularly meaningful for
very peripheral collisions where the main spectator remnant (quasi-projectile
or quasi-target) -- well identified by the MST method -- is at nearly zero
excitation energy.  


\section{Early fragment recognition.}

\begin{figure}
\begin{center}
  \includegraphics[width=0.7\linewidth]{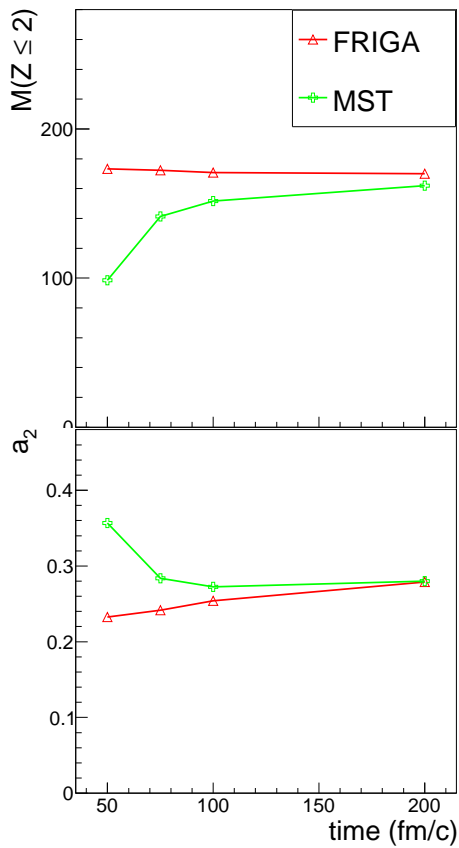}
\end{center}
  \caption{BQMD (hard equation of state) predictions of \AuAu collisions at
    600A~MeV incident energy for an impact parameter of 4~fm (semi-central).  
Top: Event multiplicity of primary small fragments with charge $Z\leq2$ identified on
the projectile side as a function of time by FRIGA and MST (respectively red
triangles and green crosses). 
Bottom: as top panel for the relative asymmetry between the primary two largest
charges of fragments identified on the projectile side (see text for the
definition of $a_2$).} 
  \label{timing}
\end{figure}

As already shown in \cite{gos97}, the application of the simulated annealing method to find the
most bound configuration allows to identify fragments much earlier in the
course of the heavy ion collision than MST. Fig.~\ref{timing} illustrates this
fact with the FRIGA clusterisation applied to  BQMD \cite{aic91} simulations
of \AuAu collisions at 600 A MeV incident energy and an impact parameter of
4~fm. Fig.~\ref{timing} top compares the time evolutions of the average
multiplicity of light particles, $Z\leq2$ of the projectile rapidity side obtained
with MST and FRIGA. For the MST algorithm  
$\Delta p = 0.6$~GeV/c and $\Delta r = 2.5$~fm have been utilized.

To compare to SACA directly (as published in Ref.~\cite{gos97}), we have only taken $E_B$ (Eq.~\ref{binding}) to calculate
the binding energies in FRIGA and omitted the asymmetry energy $B_{asy}$ and shell
effects $B_{struct}$. However we have observed that the present
results are not modified by the inclusion of these extra potentials.  
Obviously, similarly to what had been found in \cite{gos97}, the asymptotic values of the light fragment
multiplicity is reached very early with FRIGA, at around 50~fm/c, whereas MST
needs at least 200 fm/c to obtain a stable configuration. 
The same fast convergence to the asymptotic value of the 
FRIGA results is seen in Fig.~\ref{timing} bottom for the observable $a_2$, which is
relevant for the observation of bimodality in heavy ion collisions at low
incident energies \cite{lef09} and reflects the mechanical state of the system:
\begin{displaymath}
a_2 = \frac{Z_1-Z_2}{Z_1+Z_2},
 \end{displaymath}
 where $Z_1$ is the largest and $Z_2$ the second largest
 charge observed on the projectile (or target) side at the end of the
 collision. 
In general, we observe for spectator fragmentation in heavy ion
collisions with an incident energies around or below 1A~GeV,  that FRIGA can
identify final fragments as early as twice the passing time of projectile and
target, i.e. the time that they would need to completely cross each other at
zero impact parameter if the nuclei were fully transparent.

\section{De-excitation of excited fragments}
\label{sec:second}  

\begin{figure}
\begin{center}
 \includegraphics[width=0.9\linewidth]{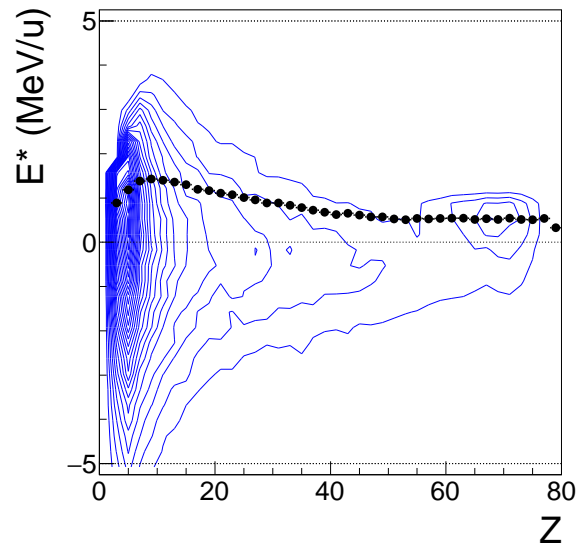}
 \end{center}
  \caption{Predictions of the BQMD transport code 
    (hard equation of state) for \AuAu collisions at
    600A MeV incident energy with impact parameters below 12~fm. A minimum
    bias distribution was generated. The figure demonstrates the resulting excitation
    energies of the primary clusters 
    identified by FRIGA as a function of their charge. The identification of
    clusters was  performed between 50 and 75~fm/c. This is two and three times longer than
    needed by the colliding system to separate. The contour plot represents
    the double differential probability distribution (linear scale). The
    black filled circles display the average values of positive excitation energies.   
}
 \label{excitation}
\end{figure}

Most of the clusters identified with FRIGA are relatively cold with a value of their binding energies 
close to the the ground state energy. They can either be over-bound (negative binding energy below that of the ground state) -- and we keep them as they are -- or slightly excited.  
These pre-fragments, called also ``primary'' fragments, can be produced ``non
relaxed'' in shape and density, i.e. deformed with respect to their fundamental
state. 
Most of the excitation energy is due to the difference in surface energy
between the reconstructed state and the respective ground state of the nucleus.  

Fig.~\ref{excitation} illustrates this variety of excitation
energies of primary clusters as a function of their charge (blue
contour lines), as identified by FRIGA after BQMD simulations of minimum bias
\AuAu collisions at 600A MeV. 
In FRIGA, we define the excitation energy $E^*$ of a fragment  by
\begin{displaymath}
E^* = (E_B + B_{asy}) -  B_{BW np}(Z,N).
\end{displaymath}

Here, shell effects are not taken into account, since they contribute
little to the excitation energy and the Bethe-Weizs\"acker formula includes
them only in parts (pairing energy). In cases where the asymmetry energy is not taken into
account when calculating the binding energy, it is neither computed in the ground state binding energy that is here the
Bethe-Weizs\"acker formula. We take the cold liquid-drop (at normal density)
formula as if it describes the "natural" fundamental state of a cold nucleus as
constructed by the QMD transport model (compare Fig.~12 of \cite{aic91}).   
We have obtained quantitatively similar results for excitation energies of primary fragments 
at incident energies between 50A~MeV and 1A~GeV.
Positive values of the excitation energy correspond to hot primary
clusters. Negative values represent nuclei which are over-bound. Over-binding may
occur in a semi-classical approach because the ground state of the
nucleus may not correspond to the lowest energy state of the quantal Hamiltonian. 
Hence, we assume that over-bound clusters are in their fundamental state  with zero
excitation energy.  The black points in  Fig.~\ref{excitation} represent the
average value of positive excitation energies -- zero excluded 
-- as a function of the fragment charge. We observe that it hardly exceeds
1A~MeV. This means that the primary fragments produced by FRIGA  are quite cold. This
is also true for lower incident energies of around 100A~MeV.  Those primary
fragments which have an excitation energy exceeding 1A~MeV can be considered
as "hot" and should undergo a secondary decay. Since the highest excitation
energies remain quite low, typical de-excitation is done via sequential
evaporation or fission. For simulating this process, we use the GEMINI++
code \cite{cha08} -- the most recent C++ version of GEMINI -- which evaluates
the production cross sections for secondary reaction products after possible
particle evaporation and/or fission.  

\begin{figure}
\begin{center}
 \includegraphics[width=\linewidth]{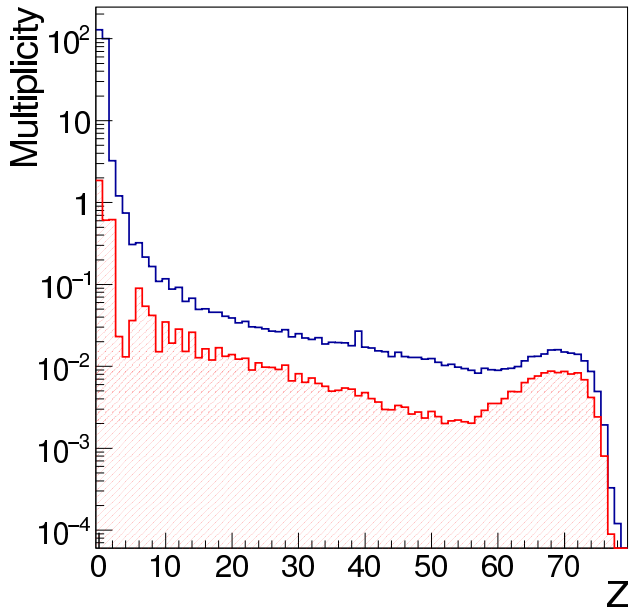}
 \end{center}
  \caption{The results for BQMD-FRIGA simulations of the average fragment
    multiplicity as function of their charge $Z$ are shown for the same
    reaction as Fig.~\ref{excitation}. The blue full line
    and the red hashed area show the overall (secondary and cold
    primary) fragment multiplicity and that of fragments resulting from
    secondary decays, respectively. In the cluster recognition procedure with
    FRIGA neither asymmetry energy nor pairing or structure effects have been
    taken into account. 
}
 \label{secondaries}
\end{figure}

Fig.~\ref{secondaries} compares the overall
(secondary and cold primary nuclei) charge yield (blue line) obtained
at twice the passing time in  minimum bias BQMD-FRIGA simulations of \AuAu
collisions at 600A MeV incident energy with the yield of fragments created by
secondary decays (red-hashed area). 
We observe that the contribution of secondary decays become
non-negligible for projectile/target remnants and for $\alpha$ particles: here 
secondaries contribute up to 50\% to the total $\alpha$ yield. In addition, 
we observe that the secondary distribution
exhibits an odd-event Z staggering at small Z which results from the pairing effects
included in GEMINI++.   

\section{Benchmarking in the spectator fragmentation regime.}


One of the main features of the spectator fragmentation at relativistic incident energy has been
discovered by the ALADiN Collaboration in the late 1990's, dubbed "Rise and Fall" \cite{sch96} curve,
exhibiting a universal behaviour which is essentially independent of the beam energy and scales with the system size. 
This curve represents the average multiplicity of intermediate mass clusters ($2 \leq Z \leq 30$) as a function of $Z_{bound}$
(total charge bound in fragments with $Z\geq 2$, which scales to the centrality of the collision). 

It has been shown that the SACA approach, using BQMD as program for the time evolution of the nucleons, can
reproduce this curve \cite{gos97}. Similar analyses applying MST do not give an equally good description \cite{beg93},
illustrating that the "Rise and Fall" is a very sensitive and challenging observable for clustering methods.  


However, as far as charge distributions are concerned, this lower accuracy of MST is less visible, 
especially in central collisions. It has been shown in
\cite{Zbiri:2006ts}, that MST used at late times on BQMD simulated collisions provides
a fair agreement with experimental charge distributions.  

\begin{figure}
\begin{center}
 \includegraphics[width=0.8\linewidth]{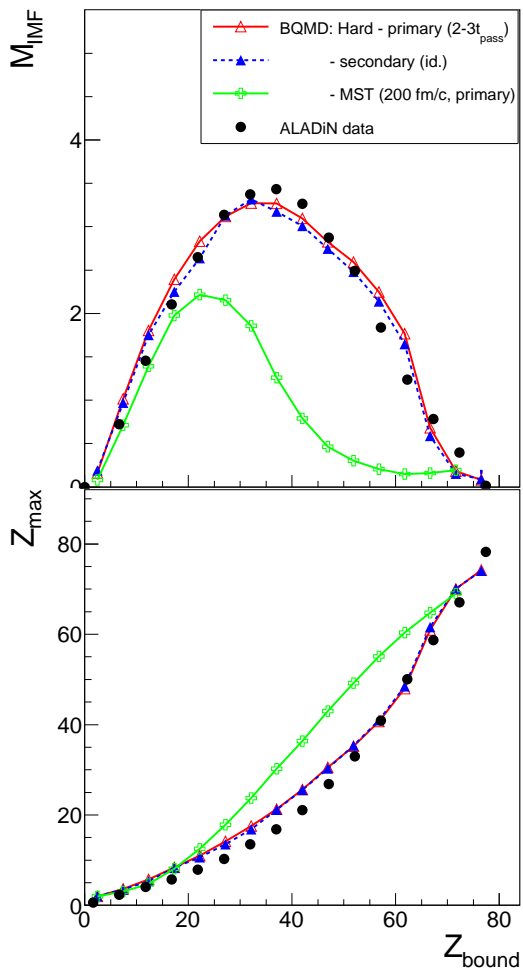}
 \end{center}
  \caption{BQMD (hard equation of state) predictions of \AuAu collisions at
    600A MeV incident energy for impact parameters b=0-12~fm, compared with
    the ALADiN S254 (2000) experimental data.  
Top: ''Rise and Fall'' curve (average multiplicity of IMF's as a function of 
the total charge of complex fragments with $Z \geq 2$, on the projectile side,
dubbed $Z_{bound}$. The green crosses depict the BQMD cluster partitions
identified with the MST method at~200 fm/c (primary clusters). The blue full
and red open triangles show respectively the primary and secondary cluster
partitions identified by FRIGA at 2 and 3 times the passing time (predictions
of both times averaged to decrease the statistical fluctuations).     
Bottom: in the representation of the average largest Z as a function of $Z_{bound}$. 
}
 \label{aladin1}
\end{figure}

To see whether FRIGA -- including secondary decays -- 
reproduces the SACA results on the ALADiN "Rise and Fall",
 we applied the FRIGA algorithm to minimum
bias \AuAu  reactions at 600A MeV, using only $E_B$ as for the calculation of
binding energies. The
result is shown in Fig.~\ref{aladin1}. There we display the average
multiplicity of  intermediate mass fragments ($2 < Z \leq 30$) on the
spectator side (top)  and the charge of the largest fragment detected on the
spectator side as a function of $Z_{bound}$. The model predictions are
compared with  the most recent ALADiN data [S254 experiment, courtesy of the
  ALADiN2000 collaboration], obtained with an upgraded set-up detailed in
\cite{sfi05} and \cite{sfi09}.  Here again, MST partitions, 
obtained at 200 fm/c, fail to reproduce the experimental
findings. On the contrary, 
the FRIGA approach -- as soon as twice the passing time -- predicts those data
with good agreement. We observed that the secondary decays do not modify
sensitively the results in these representations. The results of the GEMINI
calculations are shown as dashed lines in Fig.~\ref{aladin1}.
  
\begin{figure}
\begin{center}
 \includegraphics[width=0.8\linewidth]{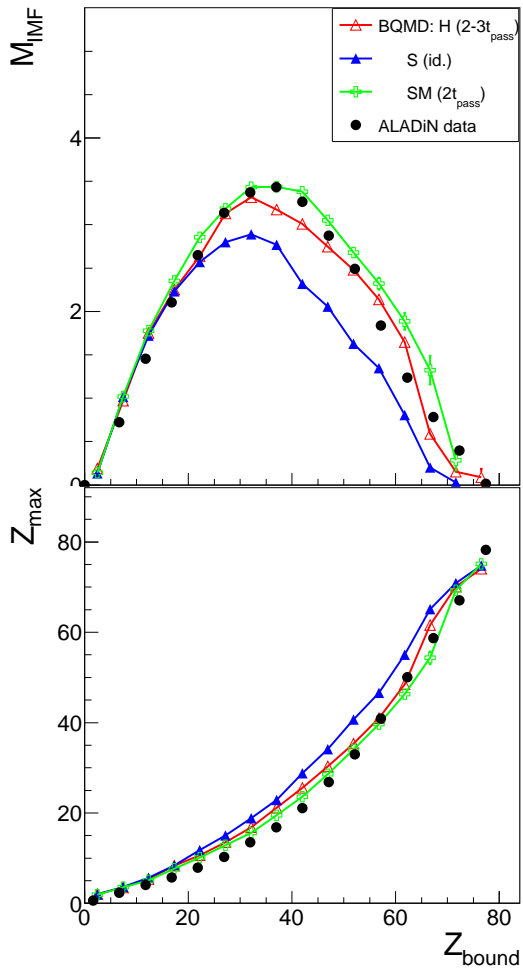}
 \end{center}
  \caption{Same as Fig~\ref{aladin1}, the FRIGA cluster partitions, after
    secondary decays, are compared to ALADiN experimental data. Three
    different parameterisations for the nuclear EoS in BQMD: hard (H),
    soft (S), and soft with momentum dependant interaction (SM), depicted
    respectively by the red open triangles, the blue full triangles and the
    green open crosses. The time interval adopted is 2-3 times the passing
    time at this energy.
}
 \label{aladin2}
\end{figure}

In order to estimate how much the equation of state (EoS) adopted in the
transport model influences the results, we compare in Fig.~\ref{aladin2} the
experimental data with the BQMD-FRIGA predictions using three different
equations of state: hard (H, red open triangles), soft (S, blue full
triangles), soft with momentum dependent interaction (green crosses). The
parameters adopted for the BQMD Skyrme potential  

\begin{displaymath}
U(\rho) = \alpha \frac{\rho}{\rho_0} + \beta \frac{\rho}{\rho_0}^\gamma +
\delta log^2(\varepsilon(\Delta {\bf p})^2+1)\frac{\rho}{\rho_0} 
\end{displaymath}

in the three configurations are listed in Table~\ref{tab:eos}. 

\begin{table}
\begin{tabular}{lcccccc}
EoS & K (MeV) &$\alpha$ (MeV)  &$\beta$ (MeV) & $\gamma$ & $\delta$ (MeV) \\
\hline
 H & 380 & -124 & 70.5 & 2. & 0 \\ 
 S & 200 & -356 & 303 & 7/6 & 0 \\
 SM & 200 & -390.1 & 320.3  & 1.14 & 1.57 \\
\end{tabular}
\caption{Parameter sets for the nuclear equation of state used in
the BQMD model. K is the nuclear incompressibility modulus derived from the
curvature of the potential at $\rho=\rho_0$}  
\label{tab:eos}
\end{table}

Like the hard (H) EoS, the soft EoS with momentum dependent interaction (SM)
reproduces well the experiment when fragments are identified with FRIGA at
around twice the passing time.
BQMD with momentum dependent interactions (m.d.i.) is not stable
asymptotically: with SM, the "Rise and Fall" results change noticeably 
with time. Therefore we do not pursue this approach further in this paper.
Adopting a soft EoS without m.d.i. (S) does not reproduce the
experimental data. 

\section{Asymmetry energy, shell effects and secondary decays}

\begin{figure}
\begin{center}
 \includegraphics[width=\linewidth]{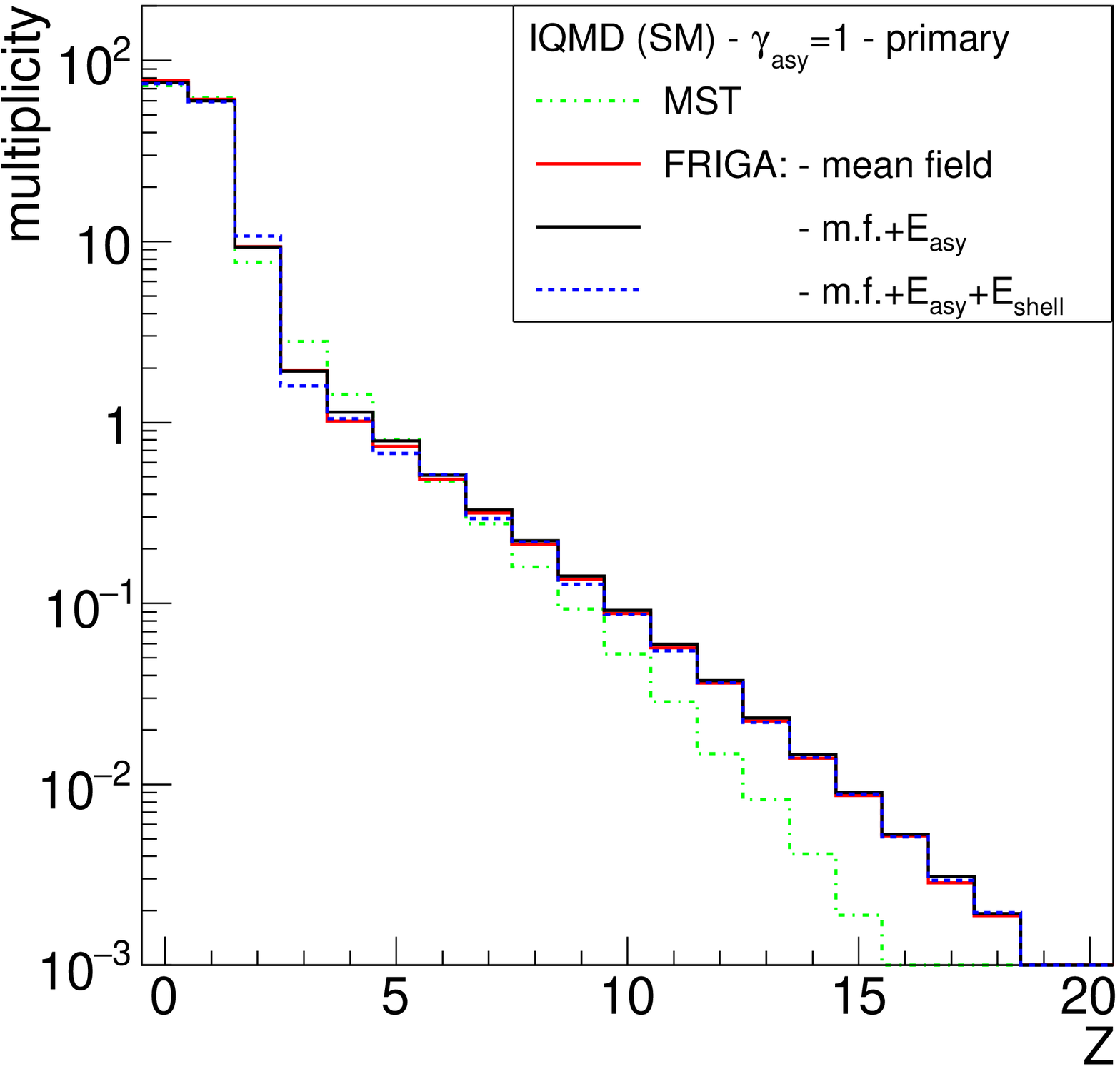}
 \includegraphics[width=\linewidth]{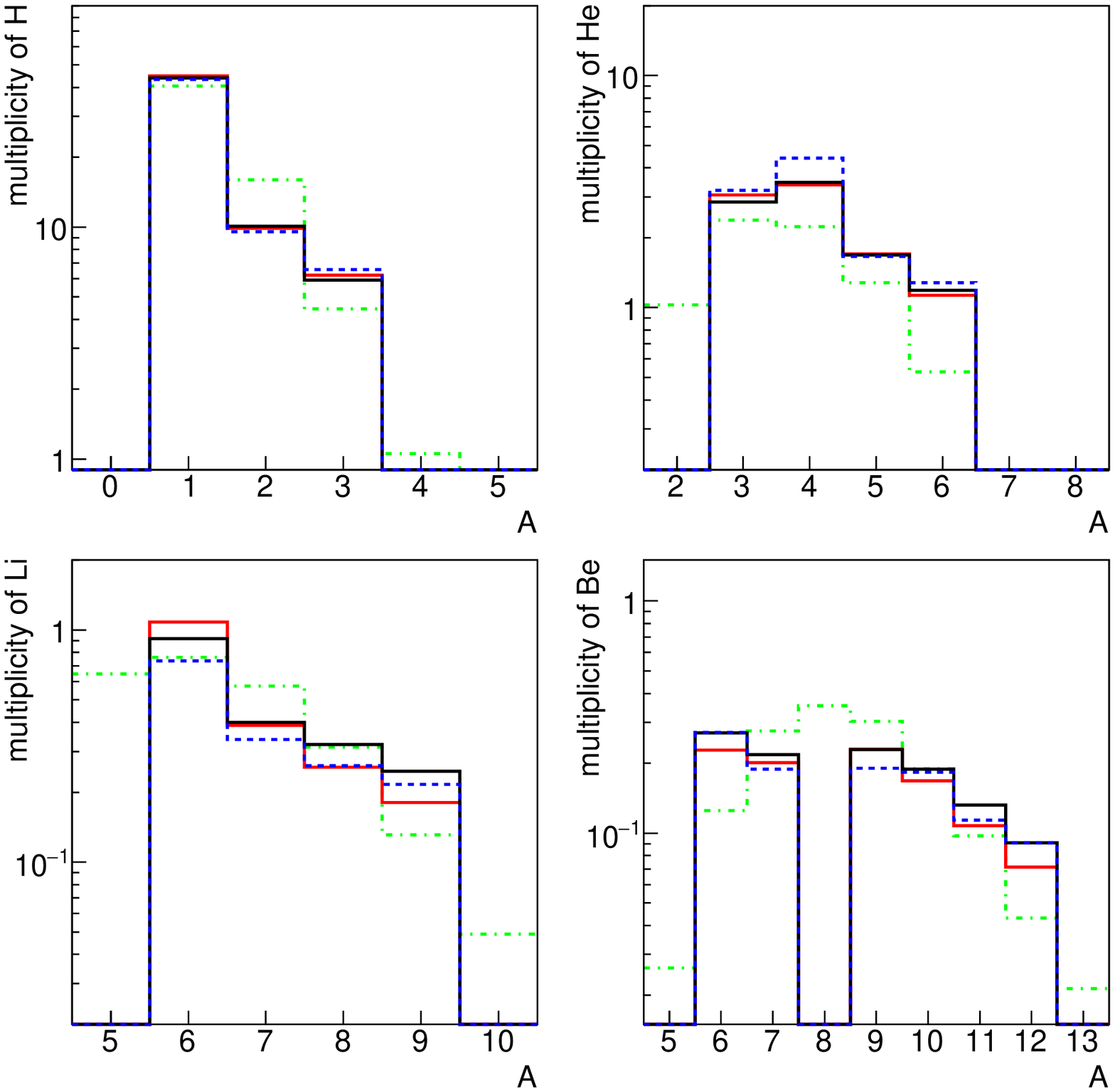}
 \end{center}
  \caption{IQMD (SM EoS) predictions of \XeSn collisions at 100A MeV incident
    energy for an impact parameter  $b<2.8 fm$ ($10\%$ most central
    collisions) and an exponent of the asymmetry potential $\gamma_{asy} = 1$.  
Top panel: Average yields in an event of primary clusters as a function of their charge. 
Bottom four-panels:  mass distributions of primary isotopes of hydrogen (top
left), helium (top right), lithium (bottom left) and beryllium (bottom
right). 
Three different FRIGA strategies are shown (applied at twice the passing
time): with the basic potential only (red full lines), with including asymmetry
potential (black full lines), and adding the shell energy (blue dashed
lines). They are compared to the result of the minimum spanning tree method
(green dotted dashed line). Colors in online version. 
}
 \label{XeSn1}
\end{figure}

In order to illustrate the influence of the various new components of the
binding energy of clusters in FRIGA, we compare the charge and light isotope
yields of primary fragments exhibited by different compositions of the cluster
binding energy in Fig.~\ref{XeSn1}. In order to infer the isotopic dependence,
we have used predictions of the IQMD code \cite{har98} which, unlike BQMD,
explicitly treats neutrons and protons with respect to mean field and
collisions and includes in its dynamics
the proton-neutron asymmetry potential $B_{asy}$. 
For benchmarking we select central \XeSn collisions at
100A~MeV incident energy for two reasons: first, they have
been measured and isotopically resolved by the INDRA detector (see below),
second, they allow to probe the binding energy configurations over a large
variety of isotopes, in a strongly dynamical environment. Here, four different
approaches are compared: 
MST alone (at 200 fm/c, based on the coordinate space proximity of nucleons), 
minimization with FRIGA employing $E_B$ only, with
$E_B + B_{asy}$, with $E_B + B_{asy} + B_{struct}$, respectively. Like in the
following we consider the cluster partitions identified at the earliest possible time when they have reached
their asymptotic characteristics. This time depends on the cluster recognition method used: It is typically twice the passing time with
FRIGA, and 200 fm/c with MST. 
 As already quoted, provided the transport model does not induce artificial modifications of the phase space extension 
 of nucleons at late times, and the primary fragments are not excited, (early) FRIGA and (late) MST
 cluster distribution should be quite identical. But since both conditions are not perfectly fulfilled with the QMD transport model, the both approaches differ a bit.  
From the Z yields of primary fragments (Fig.~\ref{XeSn1} top), we
observe first that the MST predictions do not differ strongly from those of FRIGA, apart for large clusters whose yields are under-predicted by MST. Second, there is no strong influence of the various FRIGA approaches: the effects of the asymmetry and of the shell structure on the
Z yield are very small. A stronger influence is visible if one studies the
mass distribution of small isotopes, depicted in Fig.~\ref{XeSn1} bottom. We
observe that MST exhibits broader distributions, because it is not constrained
by the vetoing of unphysical isotopes -- like $^{8}Be$ -- and is based only on
the phase space proximity. The asymmetry potential (here with a linear density
dependence) tends by nature to narrow the distributions around N =
Z. Shell effects would enhance or reduce the yields of
particular isotopes, tempting to restore the natural abundances, according to
the deviation of the experimental mass of the one given by the liquid drop
model. For instance, $\alpha $ particles in this respect are highly favored
because of their strong pairing energy.  

\begin{figure}
\begin{center}
 \includegraphics[width=0.9\linewidth]{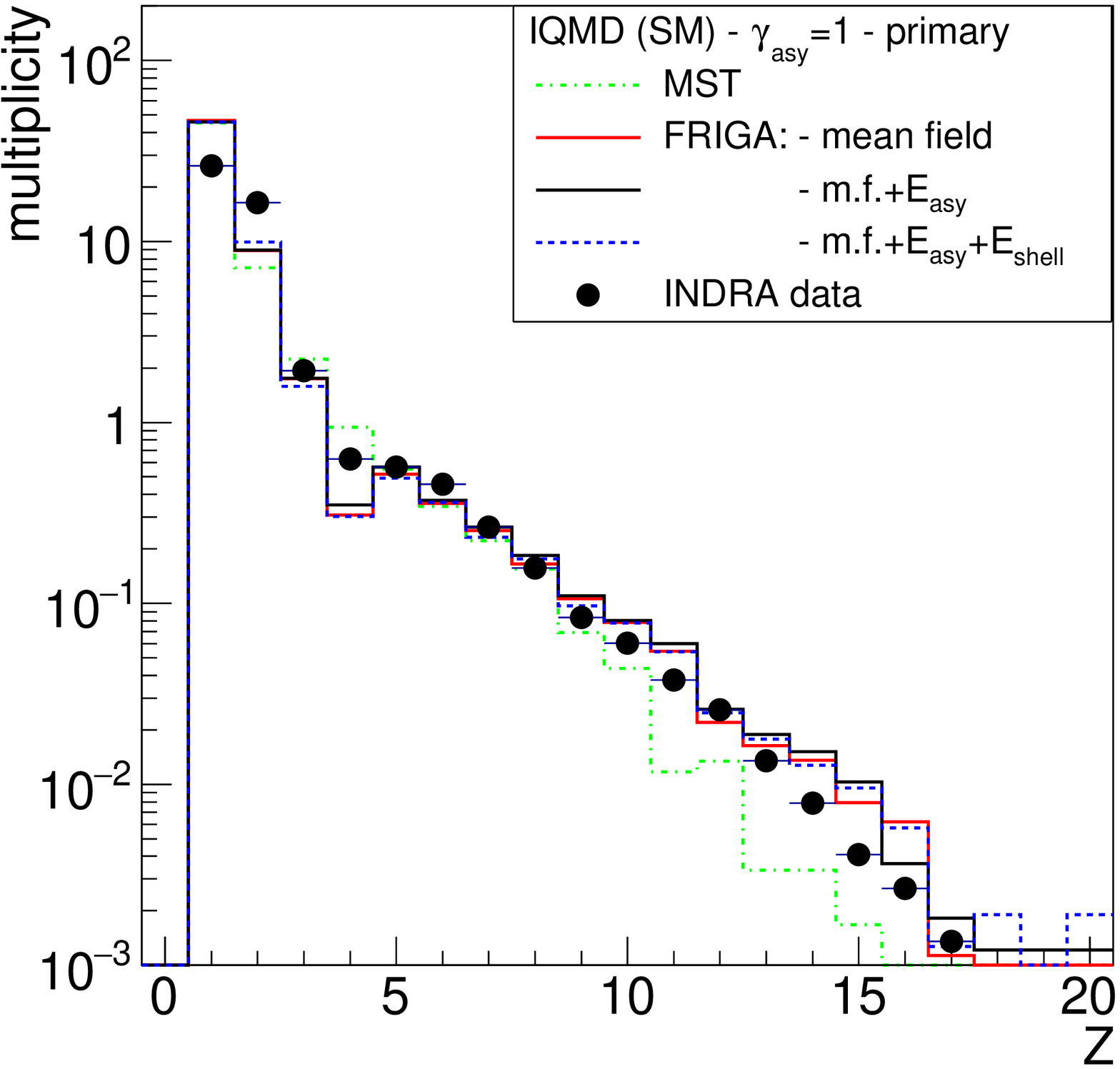}
 \includegraphics[width=0.85\linewidth]{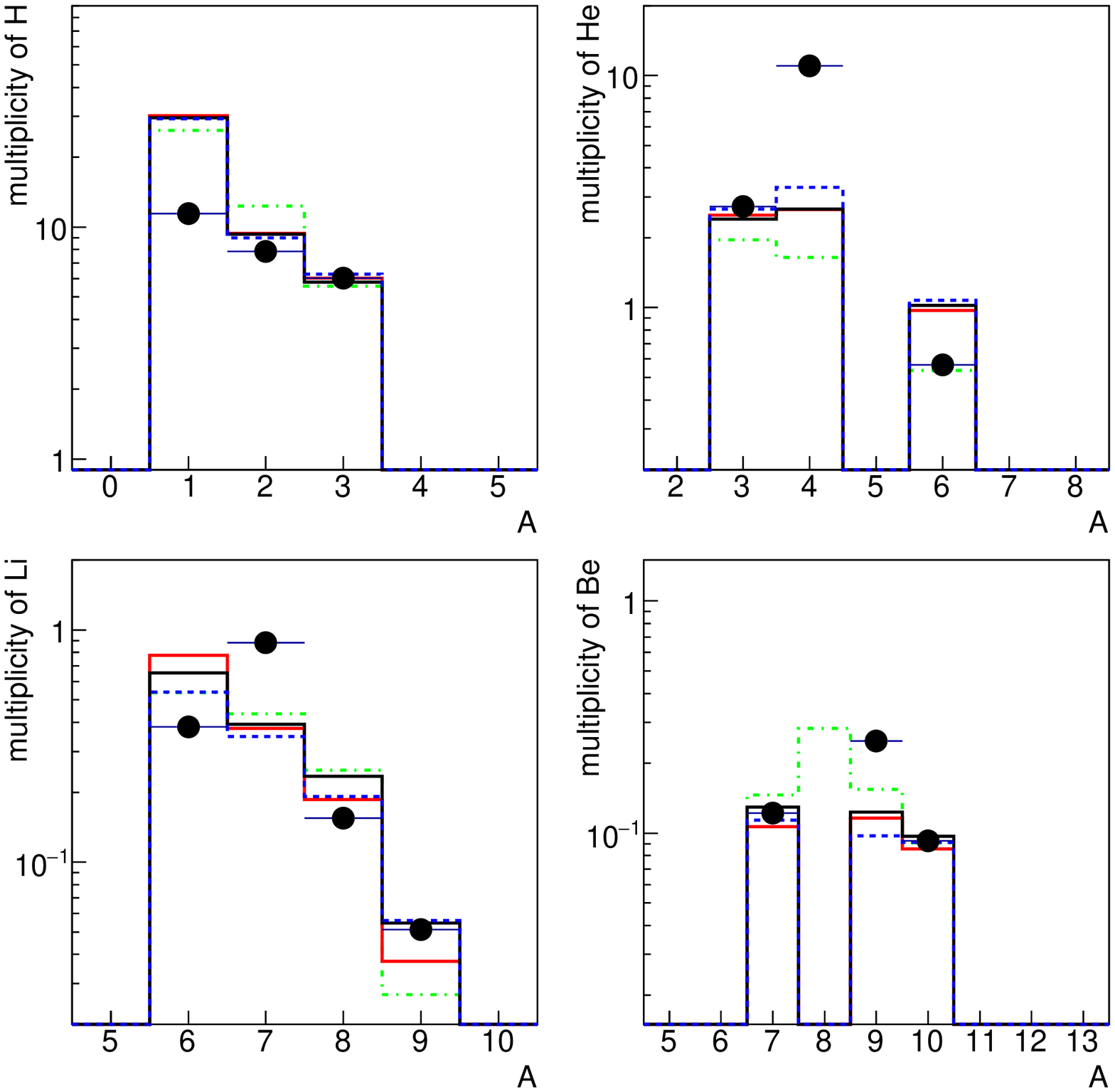}
 \end{center}
  \caption{
Same as Fig.~\ref{XeSn1} compared with the INDRA experimental data (points). 
Predictions are filtered with the software replica of the INDRA
acceptance. 
}
 \label{XeSn2}
\end{figure}

In Fig.~\ref{XeSn2} the results of the FRIGA approach are compared to
experimental data that have been measured by the INDRA detector at GSI
Darmstadt \cite{lef05}. The centrality of the events  has been selected by
means of the total transverse energy $E^{\perp}_{12}$ of detected nuclei with
charge Z~=~1 and 2, similarly to Ref.~\cite{lef05}.  The $10\%$ most central
collisions correspond to $E^{\perp}_{12} > 1440$ MeV. In order to enhance the
reliability of the experimental yields, we have selected events where 
at least $70\%$ of the total charge has been detected. To be able to compare
the model predictions to the experimental data, we have filtered the IQMD-FRIGA
events by a software replica of the INDRA acceptance. Fig.~\ref{XeSn2} top
shows that the charge yield of the FRIGA primary clusters, detected in IQMD
events, is close to the experimental data over the broad range of
charges whereas the simple phase space proximity criteria used by MST does not
give the correct slope. FRIGA predicts too many hydrogens, due to its lower
efficiency in detecting helium fragments in the hot environment of central
collisions. This may  indicate that a more complex mechanism rules the
production of the lightest isotopes in a hot expanding environment, as pointed
out in \cite{rei10}. Considering the isotope yields of Fig.~\ref{XeSn2}
bottom, primary isotopes, given by FRIGA, reproduce fairly well the
experimental yield starting from $A=6$.   

\begin{figure}
\begin{center}   
 \includegraphics[width=0.9\linewidth]{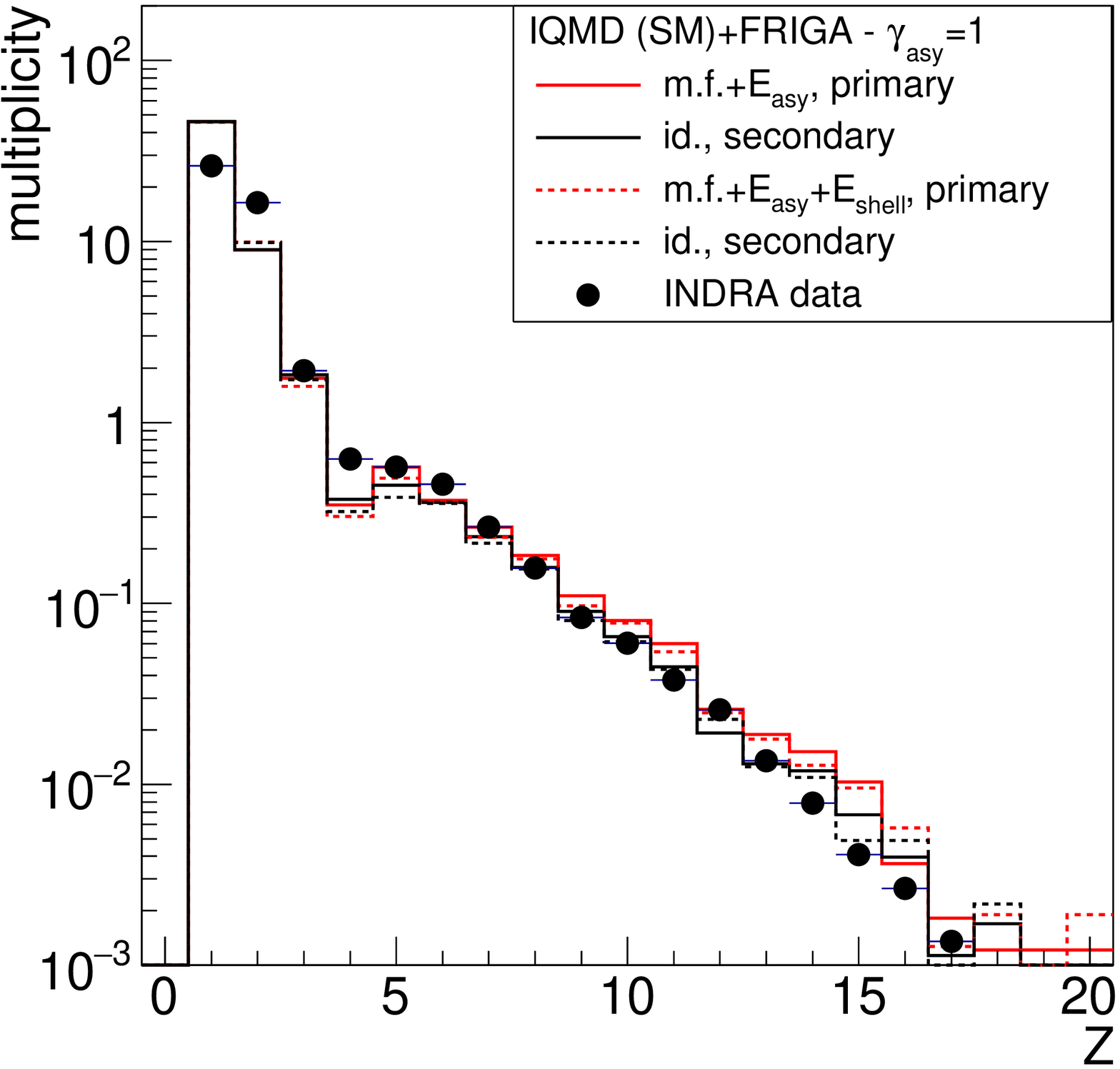}
 \includegraphics[width=0.85\linewidth]{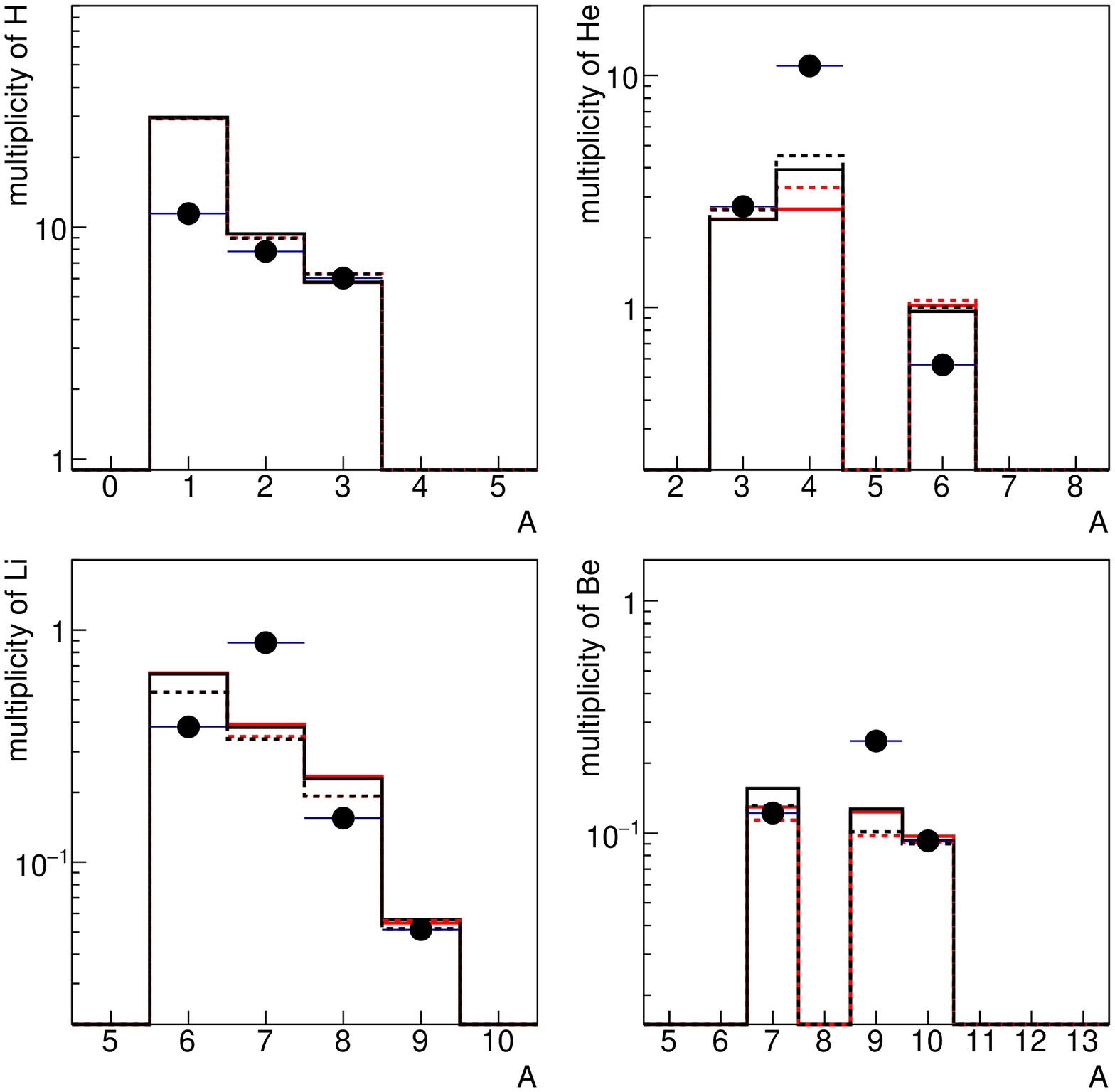}
 \end{center}
  \caption{
Same as Fig.~\ref{XeSn2} comparing the INDRA experimental data with FRIGA
predictions before (red lines) and after (black lines) secondary decays, with
the FRIGA strategy including asymmetry energy (with $\gamma_{asy}=1$).  
}
 \label{XeSn3}
\end{figure}
\begin{figure}
\begin{center}
 \includegraphics[width=0.75\linewidth]{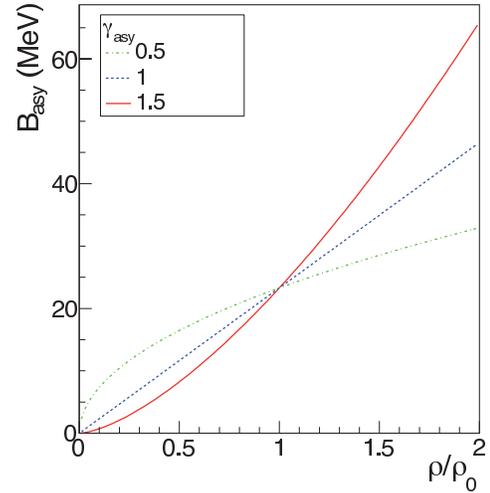}
 \end{center}
  \caption{
Density dependence of the potential part of the asymmetry energy as used in
FRIGA and IQMD for various values of the exponent $\gamma_{asy}$: 0.5 (soft
equation), 1 (stiff) and 1.5 (super-stiff), respectively displayed by green
dashed-dotted, blue dashed and red full lines.  
}

 \label{Easy}
\end{figure}

\begin{figure}
\begin{center}
 \includegraphics[width=0.75\linewidth]{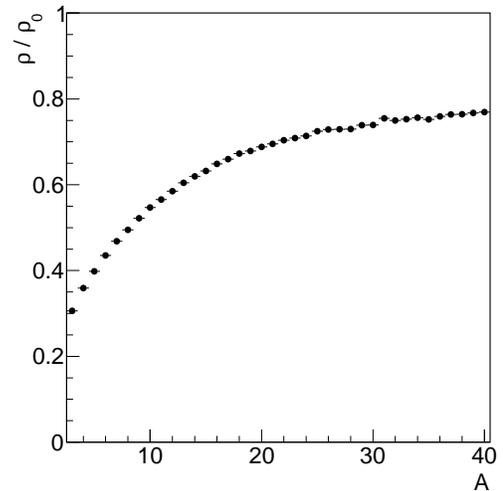}
 \end{center}
  \caption{
Average values of the internal density of clusters scaled to the saturation
density, as a function the mass number of primary clusters identified by FRIGA
(with only the basic potential in the binding energy) at twice the passing time, out of
IQMD (SM EoS, $\gamma_{asy}$ = 1) \XeSn central collisions at 100A~MeV
incident energy. 
}
 \label{density}
\end{figure}

\begin{figure}

\begin{center}
\includegraphics[width=0.9\linewidth]{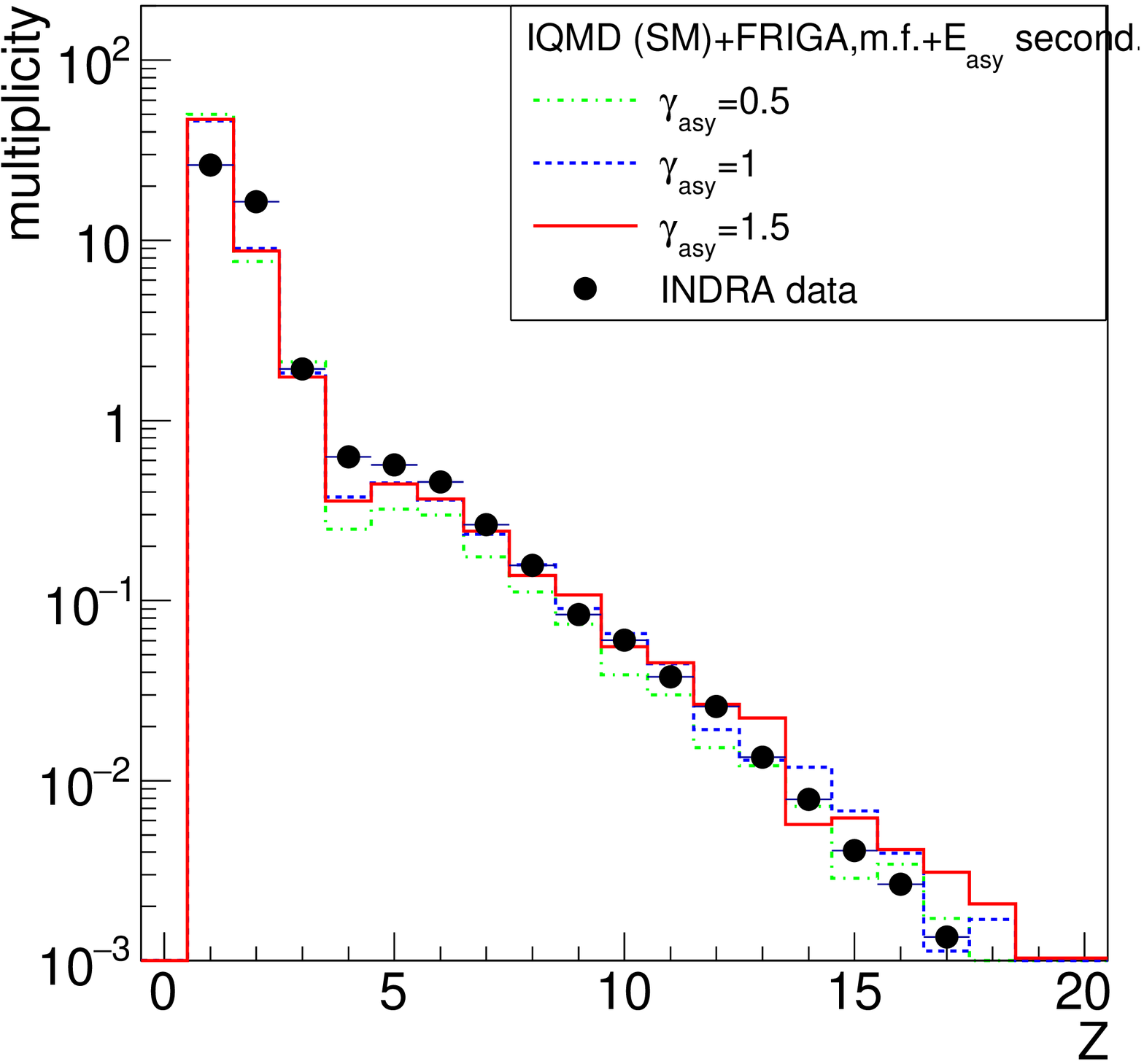}
\includegraphics[width=0.85\linewidth]{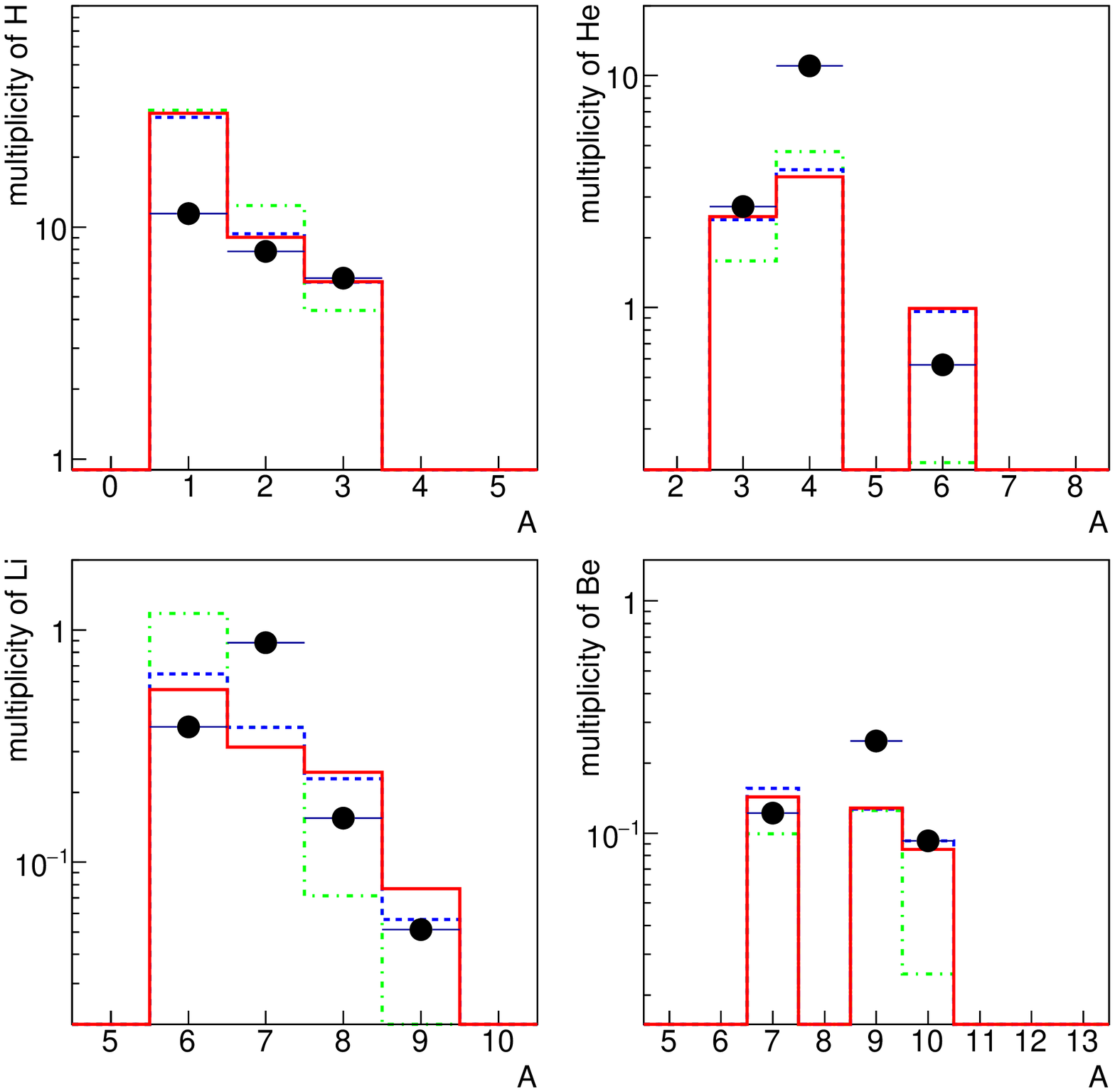}
\end{center}
   \caption{
Same as Fig.~\ref{XeSn2} comparing the INDRA experimental data with FRIGA
predictions after secondary decay with asymmetry potential included, for the
three values of the power exponent $\gamma_{asy}$ of $E_{asy}$ shown in
Fig.~\ref{Easy}, with corresponding line styles.  
}
\label{XeSn4}
\end{figure}
  
Up to now we have not included secondary decays. To illustrate their effects
on fragment yields, we choose the FRIGA strategy including the asymmetry
energy and shell effects. Similar results are observed when considering only the basic potential
for the binding energy. The results are shown as black lines in Fig.~\ref{XeSn3} for secondary
fragments partitions, compared with the primary yield (red lines). The main
consequence of secondary de-excitations is an increase of the yield of small
fragments (with $Z < 5$) at the cost of larger ones, which brings to a better
agreement with the INDRA experimental charge distribution. The main channels
of de-excitations are the emissions of neutrons and alphas particles, which is
reflected by the enhancement of this latter ones as seen in Fig.~\ref{XeSn3}
bottom. 
From the isotope yields, we conclude that shell effects in primary clusters
are too weak to exhibit a clear difference. The reproduction of experimental
yields of alpha particles is slightly improved by shell effects (after
secondary decays) on a way, but on the other one, it is worsened for $^{9}Be$.  
The absence or not of primary shell effects is not obvious, because,  in order
to be stable, even primary fragments must be quite cold -- therefore shell
effects may still survive.  On the other side, they are not in vacuum but
tightly surrounded by a hot medium whose temperature is of the order of 6 MeV
in these collisions, as shown in \cite{lef05}. This may prevent a realization
of structure effects inside the clusters in an early phase.   

As we have seen when inspecting the width of the isotope mass distributions, the
asymmetry energy in primary clusters is a key ingredient of the binding energy
for describing correctly the final isotope distributions. Therefore, it may be
possible that the stiffness of its density dependence has  a measurable
influence on the observables. Fig.~\ref{Easy} displays the evolution of
$B_{asy}$ as a function of the density
for various values of the exponent $\gamma_{asy}$. A larger exponent implies a
stronger asymmetry potential at supra-saturation densities, and reversely at
sub-saturation densities.
The cluster internal density -- that we call "intrinsic" -- is determined in the very same way as in
IQMD \cite{har98}:  
\begin{equation} \label{rhoint}
\rho_{\rm int}^i(\vec{r_i}) = \frac{1}{(\pi L)^{3/2}} 
\sum_{j \neq i} {\rm e}^{\displaystyle 
-(\vec{r_{i}}-\vec{r_{j}})^2/L }
\end{equation}
As already quoted, the density used for calculating the binding energy in FRIGA 
is not that of the medium, but that intrinsic to the cluster which is typically close to that of its ground state.   
Therefore, the cluster formation in FRIGA probes only sub-saturation densities  typically
ranging between 0.3 and 0.8 times  $\rho_0$, as illustrated in
Fig.~\ref{density} for the case of \XeSn central collisions at 100A~MeV
incident energy. The average density increases with the fragment
size. Hence, the strongest sensitivity on the stiffness of the asymmetry
energy is expected for small to intermediate mass fragments, i.e. for  $A \leq
20$. On its side, the average density of the medium has been observed to be close to $\rho_0$ when the partition of clusters identified by FRIGA are first stabilized.
The fragments identified by FRIGA have a smaller density, typically around
$\rho=\rho_{0}/2$ for intermediate mass fragments, and around
$\rho=\rho_{0}/5$ for the light $Z<3$ isotopes. In particular during the maximum overlap of the colliding system, nearby nucleons can happen to form a dense group, but they have quite different
velocities in the beam direction. They do, however, not form a common fragment
because its internal kinetic energy would be too high. 

 For the same system we show in Fig.~\ref{XeSn4} the sensitivity of fragment
 partitions on $\gamma_{asy}$, as predicted by FRIGA. We use the basic and
 asymmetry potentials to calculate the cluster binding energy and observe
 that there is no strong influence on the charge multiplicities
 (Fig.~\ref{XeSn4} top), except for beryllium
as seen in Fig.~\ref{XeSn4} bottom right. The reason is that  
 a stronger asymmetry energy (therefore lower $\gamma_{asy}$ at sub-saturation density) 
 disfavours beryllium isotopes other than $^{8}Be$, but this latter decays into two alpha particles 
 in the final secondary de-excitation procedure.
 The mass distributions of light isotopes
 (Fig.~\ref{XeSn4} bottom) are strongly influenced by $\gamma_{asy}$, in
 particular the heaviest elements. Concerning hydrogen, deuterons and tritons, their yields show a slight
 sensitivity to $\gamma_{asy}$. The general trend is that the softer the asymmetry potential
 (smaller $\gamma_{asy}$), the narrower is the mass distribution around $A =
 2Z$. Here, comparing our results with the experimental distributions of
 beryllium and lithium, we observe that a moderately soft asymmetry potential is
 favoured.

\section{The hyper-nucleus formation.}

\begin{figure}
\begin{center}
 \includegraphics[width=0.7\linewidth]{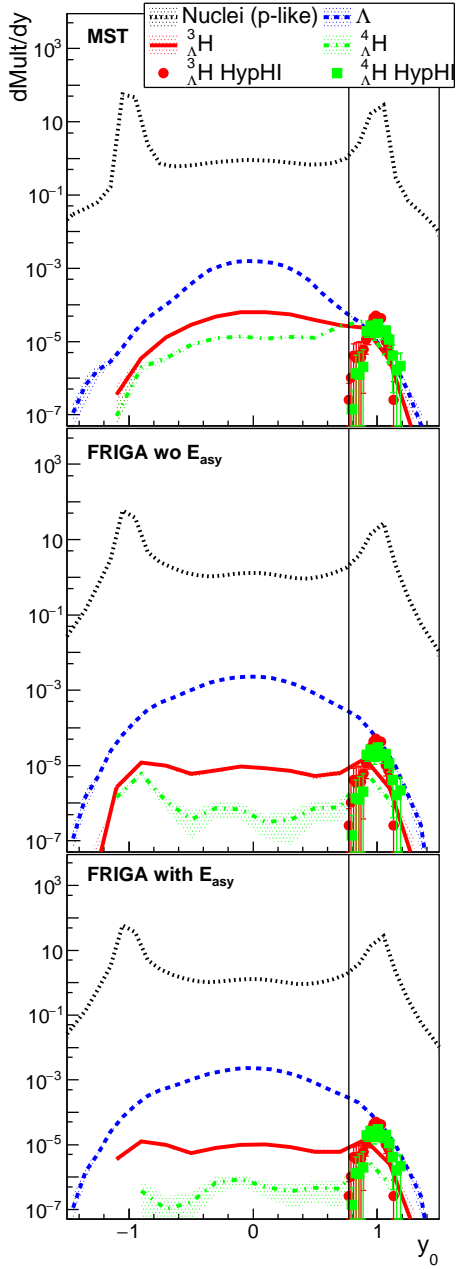}
 \end{center}
  \caption{
FRIGA predictions from IQMD (SM EoS) calculations of \LiC collisions at 2~AGeV 
incident energy for an impact parameter $2.0<b<5.5 fm$ at twice the passing
time: rapidity dependance of yields per event per unit of rapidity of
$\Lambda_0$ hyperons and clusters, compared with HypHI experimental yields of
hyper-tritons $^{3}_{\Lambda}H$ (red dots) and $^{4}_{\Lambda}H$ (green
squares) from \cite{rap15}. 
Model predictions of yields of overall nuclei (in proton-like weighting),
$\Lambda_0$ hyperons, hyper-tritons and $^{4}_{\Lambda}H$ are indicated
respectively with black dotted, blue dashed, red full and green dashed-dotted
lines.  
The results of the model calculations are not filtered for the experimental acceptance. 
The rapidity is expressed in the reference frame of the nucleon-nucleon centre
of the colliding system, and scaled to the projectile rapidity. The vertical
full line indicates the rapidity above which the HypHI acceptance is alleged
to be close to $100\%$. Below, the experimental acceptance limits the
available phase space.  
Top panel: with clustering done with the MST method.
Middle panel: with clustering done by FRIGA with only the basic potential. 
Bottom panel: with clustering done by FRIGA with asymmetry energy in addition
(with $\gamma_{asy}=1$).  
}
 \label{HypHI1}
\end{figure}

\begin{figure}
\begin{center}
\includegraphics[width=0.7\linewidth]{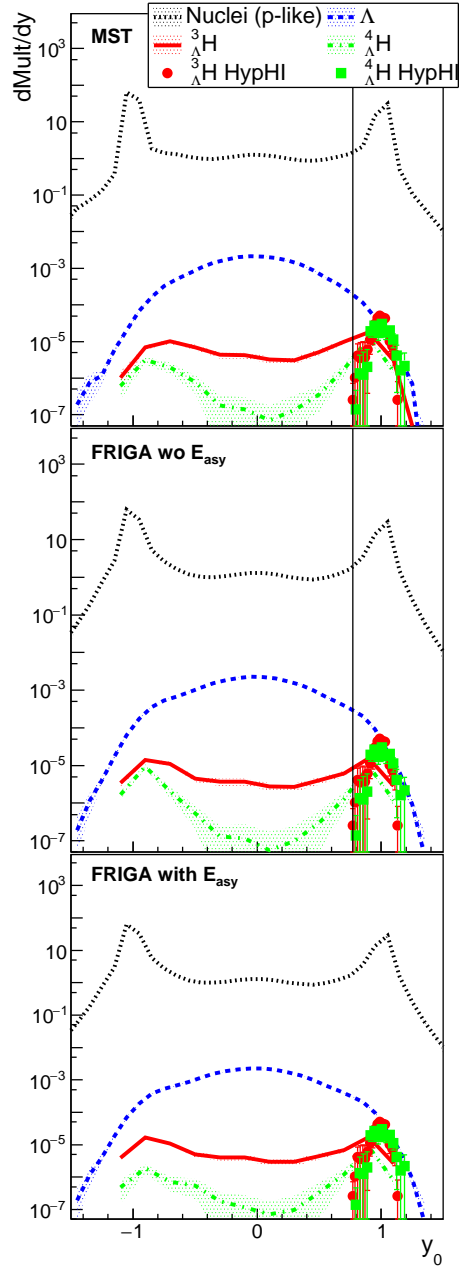}
\end{center}
  \caption{
Same as Fig.~\ref{HypHI1} with FRIGA clustering performed at four times the passing time.
}
\label{HypHI2}
\end{figure}

\begin{figure}
\begin{center}
\includegraphics[width=0.7\linewidth]{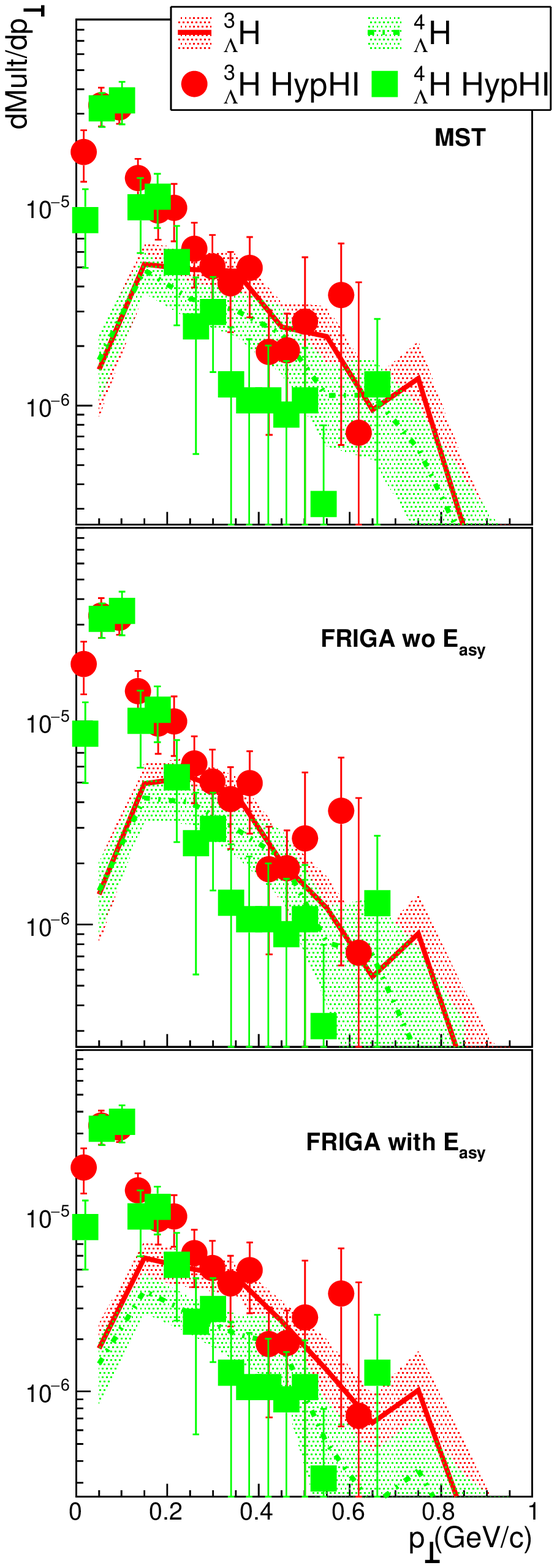}
\end{center}
  \caption{
Same as Fig.~\ref{HypHI1}, at four times the passing time, with only
hyper-tritons and $^{4}_{\Lambda}H$ in the representation of the transverse
momentum in the projectile spectator region corresponding to the HypHI
acceptance. Experimental data are extracted from \cite{rap15}. 
}
\label{HypHI3}
\end{figure}

A hyper-nucleus is a nucleus which contains at least one hyperon
($\Lambda(uds)$, ...) in addition to nucleons. 
Here we will restrict ourselves to hyper-nuclei composed by $\Lambda_{0}$ hyperons. 
Extending FRIGA to the strange sector requires the knowledge of the hyperon-N (here $\Lambda$N) potential.
In this first study, we consider the strange quark as inert and use 
$V_{\Lambda N} = \frac{2}{3} V_{NN}$ for protons as well as for neutrons.
Similarly, we consider the case of multiple strange nuclei as well, in which
more than one hyperon is bound in a fragment. 
There, the coupling of 2  $\Lambda's$ contributes with the potential 
$V_{\Lambda\Lambda} = (\frac{2}{3})^{2} V_{nN}$. In the present approach
we neglect a possible contribution of the hyperons to the asymmetry energy, and take, as far
as the asymmetry energy is concerned, only the contribution of the core of non-strange
nucleons  
as if it were decoupled from the hyperon. Since  for hyper-nuclei the pairing
and shell contributions to the binding energy are not well known,   
we neglect the $B_{struct}$ contribution. 

Using these modifications of the potentials, FRIGA identifies hyper-nuclei
with the same procedure as 
non strange fragments. In the underlying transport program, $\Lambda$'s are produced in different reactions:  
$\bar{K}+N\rightarrow\Lambda+\pi$, $\pi+N\rightarrow\Lambda+K^{+/0}$,
$\pi^{-}+p\rightarrow\Lambda+K_{0}$, $p+p\rightarrow\Lambda+X$.
Their abundance, position and momentum distributions are strongly
influenced by the reaction kinematics, the nuclear equation of state and the 
in-medium properties of the $K^+$ and $K^-$ (kaon potential, etc.) \cite{har11}.  

Hyper-nuclei are produced when a cluster in coordinate and 
momentum space absorbs a hyperon. In heavy-ion collisions at relativistic energies, 
the hyperon distribution is strongly peaked around mid-rapidity whereas the large 
fragments have rapidities close to the beam or target rapidity. The closer the
rapidity of the hyperon approaches 
-- by production or by subsequent collisions -- the target/beam rapidity, the
larger is the probability that it can be absorbed by  
one of the heaviest fragments. Heavy hyper-nuclei are therefore produced not far
away from beam/target rapidity. 
Hyperons can also form light clusters at
mid-rapidity with other nucleons. There, the probability decreases 
with the cluster size because it is increasingly difficult to form large
clusters out of a gas of nucleons. Whereas the large 
clusters in the beam/target rapidity regime can be identified quite early, the
light clusters at mid-rapidty are formed late 
and many of them dissolve due to the interactions with the surrounding nucleons
which form a gas at high temperature as compared to the cluster binding
energy.  

As discussed in the previous chapters, the ingredients of the cluster binding
energy influence the light isotope yields in FRIGA. The 
same is observed for hyper-nuclei. We have observed that the reduction factor $\frac{2}{3}$ in $V_{\Lambda N}$
has a noticeable effect by decreasing  
the average hyper-nuclei yields by around 20 percent. The asymmetry energy in
the cluster can have a similar effect, depending 
on the core (Z,N) asymmetry. 

In order to illustrate the predictive power of the FRIGA algorithm, we
confront results to experimental observations of light hyper-nuclei produced in the
projectile 
spectator region in collisions of  \LiC at 2A GeV incident energy, measured by
the HypHI collaboration  
at the SIS18 synchrotron of GSI Darmstadt. The data are taken from \cite{rap15}.
Fig.~\ref{HypHI1} compares the IQMD-FRIGA predictions with the experimental rapidity
distributions of ${}^{3}_{\Lambda}H$ and ${}^{4}_{\Lambda}H$. 
The acceptance of the HypHI set-up allows to reconstruct hyper-nuclei starting
from a reduced rapidity, $y_0 = y/y_{proj} \approx 0.8$ in the nucleon-nucleon
centre-of-mass system (y and $y_{proj}$ are the rapidity and projectile
rapidity in the chosen reference frame, respectively). This  corresponds to $y_{lab} =
1.6$ in the laboratory frame. Therefore, we limit our
comparison to this rapidity region, assuming that the complex experimental
trigger 
does not require any extra cuts on the simulation data. However, we observed that
a better agreement with the very peaked experimental hyper-hydrogen rapidity distribution is obtained when
  excluding the most central collisions (taking b$> $2 fm), which indicates that the experimental trigger might have favoured
  peripheral events. Therefore we adopt this centrality cut for the following. 
From the proton-like distributions predicted by IQMD, we see that the rapidity region chosen by
HypHI exhibits the highest hadronic yield and contains still the tail of the
$\Lambda$ distribution. 
Taking  MST as cluster recognition method (Fig.~\ref{HypHI1} top) results in
fairly good agreement of the hyper-hydrogens yields in the projectile spectator region. 
In this region, we note that
the MST yield of hyper-nuclei is high enough to create a visible depletion of
remaining free $\Lambda$ hyperons.  
With the FRIGA approach, 
we obtain slightly less hyper-hydrogens than with
MST. Looking at the yield ratio
$Y({}^{3}_{\Lambda}H)/Y({}^{4}_{\Lambda}H)$ allows to infer the effect of the
asymmetry energy of the core nucleus in FRIGA. In the accepted rapidity range,
the HypHI experiment has measured a yield ratio
$Y({}^{3}_{\Lambda}H)/Y({}^{4}_{\Lambda}H) = 1.4 \pm 0.8$. With the asymmetry
energy included in FRIGA, at twice the passing time, the predicted yield ratio
is $4.6 \pm 1.1$, whereas without $B_{asy}$, the ratio goes down to $2.7 \pm
0.5$, because the asymmetry energy tends to reduce the production of
$^{4}_{\Lambda}H$, whose core is the isospin-asymmetric triton.  
However, it turns out that these ratios stabilise in time slightly later in
the course of the collision.

In order to probe its persistence with time, we have performed the same
comparison at $4t_{pass}$. The result is shown in  
Fig.~\ref{HypHI2}.
At the later time, the hyper-hydrogen yields decrease by an order
of magnitude in the mid-rapidity region, but they are not strongly modified in the
vicinity of the projectile/target spectator rapidity when using FRIGA. 
The MST results show a reduction of the projectile/target region yields, 
which become similar to the ones of the FRIGA approach. 
The reason is a
cooling down of the spectator phase-space. However, the yield ratios
$Y({}^{3}_{\Lambda}H)/Y({}^{4}_{\Lambda}H)$ predicted by FRIGA tend to get smaller, 
to values which come closer to the experimental results:   
$2.0 \pm 0.4$  and $2.5 \pm 0.5$ respectively without and with $B_{asy}$ in
the binding energy. Therefore, though the FRIGA parametrisation without
$B_{asy}$ seems to be favoured in comparison with the experiment, the alternative
strategy, including asymmetry energy in the core nucleus cannot be ruled-out.   


Fig.~\ref{HypHI3} shows that the distributions of the transverse momentum
$p_\perp$ in the projectile spectator region agree well (here at four times
the passing time), in the slopes and the absolute yields at large transverse
momenta with HypHI results, independent of the clustering strategies. A noticeable
discrepancy appears at low transverse momenta where the predicted yields are
cut-off. This is mainly induced by too few low transverse momentum
$\Lambda_0's$ generated in the spectator region by our present transport model
(IQMD). This depletion at low transverse momenta remains unchanged at earlier
times. 
It could explain the yield underestimation that we noticed on the rapidity
projection (Figs.~\ref{HypHI1} and \ref{HypHI2}).

\section{Conclusion}
We present here the first step towards an understanding of the production of
isotopic yields and hyper-nuclei 
in heavy ion reactions. In order to study these we have developed the
clusterization algorithm FRIGA,  which is based on the SACA approach. 
The new features include asymmetry and pairing energies as well as shell effects. These are necessary to describe more precisely the nuclear binding energy 
than it was possible in SACA.
The density, temperature and density dependence of these contributions to the binding energy are only vaguely known. They have to be adjusted by comparing the results of the FRIGA algorithm with the existing experimental data. For the interaction between $\Lambda$ and non-strange nucleons we use here a very simplified approach assuming that the strange quarks does not contribute to the interaction. We observe that the asymmetry potential can have a strong influence on the yields of $(hyper-)isotopes$.  According to this model, the nucleons which form fragments have initially a density close to that of their ground state -- below the saturation density --, which may differ from the density of the surrounding medium. They contract a little and form finally slightly excited fragments which may undergo secondary decays. Therefore, the fragment formation is sensitive to the sub-saturation density dependence of the asymmetry energy. Shell structure effects in primary clusters seem, however, to be of less importance when we compare to the isotope yield.

In this first study we investigated the influence of these new ingredients on the fragment yield and show that the approach allows for realistic predictions 
of the absolute (hyper-)isotope yield at relativistic energies in the domain of spectator fragmentation as well as of multi-fragmentation at intermediate energies. 
In particular, for the first time, HypHI experimental yields of light hypernuclei could be quantitatively predicted within the experimental acceptance.   

{\bf Acknowledgment} : We thank Prof. Elena Bratkovskaya and V. Kireyeu for stimulating discussions and the ALADiN and INDRA collaborations for the permission to use their data prior to publication.


\begin{thebibliography}{99}
\bibitem{sch96} A. Sch\"uttauf et al., Nucl. Phys. A \textbf{607} (1996) 457.
\bibitem{rei10} W. Reisdorf et al., Nucl. Phys. A \textbf{848} (2010) 366-427.
\bibitem{Buss:2011mx} 
  O.~Buss {\it et al.},
  Phys.\ Rept.\  {\bf 512}, 1 (2012)
  doi:10.1016/j.physrep.2011.12.001
  [arXiv:1106.1344 [hep-ph]].
\bibitem{Weil:2016zrk} 
  J.~Weil {\it et al.},
  Phys.\ Rev.\ C {\bf 94}, no. 5, 054905 (2016)
  doi:10.1103/PhysRevC.94.054905
  [arXiv:1606.06642 [nucl-th]].
\bibitem{Danielewicz01}C Kuhrts, M Beyer, P Danielewicz, G Röpke, Phys. Rev. C \textbf{63} (2001) 034605
\bibitem{Feldmeier90} H. Feldmeier. Nucl. Phys. A \textbf{515} (1990) 147
\bibitem{Ono92} Ono, Horiuchi et al., Prog. Theor. Phys. \textbf{87} (1992) 1185.
\bibitem{Bauer87} W. Bauer, G. F. Bertsch, and S. Das Gupta, Phys. Rev. Lett. \textbf{58} (1987) 863 
\bibitem{Guarnera96} A. Guarnera et al., Phys. Lett. B \textbf{373} (1996) 267-274
\bibitem{Colonna98} M. Colonna et al., Nucl. Phys. A \textbf{642} (1998) 449-460
\bibitem{aic91}J. Aichelin. Phys. Reports \textbf{202}, 233 (1991).
\bibitem{har98} Ch. Hartnack et al., Eur. Phys. J. A \textbf{1} (1998) 151.

\bibitem{Bass:1998ca} 
  S.~A.~Bass {\it et al.},
  Prog.\ Part.\ Nucl.\ Phys.\  {\bf 41}, 255 (1998)
  [Prog.\ Part.\ Nucl.\ Phys.\  {\bf 41}, 225 (1998)]
\bibitem{Gossiaux:1994jq} 
  P.~B.~Gossiaux, D.~Keane, S.~Wang and J.~Aichelin,
  Phys.\ Rev.\ C {\bf 51}, 3357 (1995).
\bibitem{Steinheimer:2012tb} 
  J.~Steinheimer, K.~Gudima, A.~Botvina, I.~Mishustin, M.~Bleicher and H.~Stocker,
  Phys.\ Lett.\ B {\bf 714}, 85 (2012)
  doi:10.1016/j.physletb.2012.06.069
  [arXiv:1203.2547 [nucl-th]].
\bibitem{Botvina:2016fav} 
  A.~S.~Botvina, M.~Bleicher, J.~Pochodzalla and J.~Steinheimer,
  Eur.\ Phys.\ J.\ A {\bf 52}, no. 8, 242 (2016).
  doi:10.1140/epja/i2016-16242-7


\bibitem{gut76} H.H. Gutrod et al., Phys. Rev. Lett. \textbf{37} (1976) 667. 
\bibitem{Gosset77} J. Gosset et al., Phys. Rev. C \textbf{16} (1977) 629.
\bibitem{lem79} M.C. Lemaire et al., Phys. Lett. \textbf{85B} (1979) 38.
\bibitem{sat81} H. Sato and K. Yazaki, Phys. Lett. \textbf{98B},3 (1981).
\bibitem{gos97} P.B. Gossiaux, R. Puri, Ch. Hartnack, J. Aichelin, Nucl. Phys. A \textbf{619} (1997) 379-390.
\bibitem{dor93} C. O. Dorso and J. Randrup, Phys. Lett. B \textbf{301}, 328 (1993).
\bibitem{pur00} R. K. Puri and J. Aichelin,  J. Comput. Phys. \textbf{162}, 245 (2000).
\bibitem{zbi07} K. Zbiri, A. Le F\`evre, J. Aichelin et al., Phys. Rev. C \textbf{75} (2007) 034612. 
\bibitem{lef09} A. Le F\`evre et al., Phys. Rev. C \textbf{80} (2009) 044615.
\bibitem{cha08} R.J. Charity, GEMINI: \textit{A Code to Simulate the Decay of a Compound Nucleus by a Series of Binary Decays}, in Joint ICTP-AIEA Advanced Workshop on Model Codes for Spallation Reactions, Report  INDC(NDC)-0530 (IAEA, Vienna, 2008), p. 139.
R. J. Charity, Phys. Rev. C \textbf{82} (2010) 014610.
D. Mancusi, R. J. Charity, J. Cugnon, Phys. Rev. C \textbf{82} (2010) 044610.
\bibitem{Zbiri:2006ts} 
  K.~Zbiri {\it et al.},
  Phys.\ Rev.\ C {\bf 75}, 034612 (2007)
  doi:10.1103/PhysRevC.75.034612
  [nucl-th/0607012].
\bibitem{beg93} M. Begemann-Blaich et al., Phys. Rev. C \textbf{48} (1993) 610.
\bibitem{sfi09} Sfienti et al., Phys. Rev. Lett. \textbf{102} (2009) 152701.
\bibitem{sfi05} Sfienti et al., Nucl. Phys. A \textbf{749} (2005) 83-92.
\bibitem{cas09} W. Cassing, E.L. Bratkovskaya, Nucl. Phys. A \textbf{831} (2009) 2.
\bibitem{Hartnack:2011cn}
  C.~Hartnack, H.~Oeschler, Y.~Leifels, E.~L.~Bratkovskaya and J.~Aichelin,
  Phys.\ Rept.\  {\bf 510} (2012) 119

\bibitem{kha07} E. Khan, Nguyen Van Giai, N. Sandulescu, Nucl. Phys. A \textbf{789} (2007) 94.
\bibitem{lef05} A. Le F\`evre et al., Nucl. Phys. A \textbf{735} (2005) 219-217.
\bibitem{har11}
  C.~Hartnack, H.~Oeschler, Y.~Leifels, E.~L.~Bratkovskaya and J.~Aichelin,
  Phys.\ Rept.\  {\bf 510} (2012) 119
  [arXiv:1106.2083 [nucl-th]].
\bibitem{rap15}  Ch. Rappold et al., Phys. Lett. B \textbf{747} (2015) 129. 

\end{thebibliography}
\end{document}